\documentclass[aps,twocolumn,floatfix,showpacs,superscriptaddress]{revtex4-1}
\usepackage{graphicx,txfonts,bm}
\usepackage[colorlinks,citecolor=magenta,linkcolor=blue]{hyperref}

\begin{document}

\title{Electronic correlations of element cerium under high pressure}
\author{Haiyan Lu}
\affiliation{Science and Technology on Surface Physics and Chemistry Laboratory, P.O. Box 9-35, Jiangyou 621908, China}
\affiliation{Beijing National Laboratory for Condensed Matter Physics, and Institute of Physics, Chinese Academy of Sciences, Beijing 100190, China}
\author{Li Huang}
\email[Corresponding author: ]{lihuang.dmft@gmail.com}
\affiliation{Science and Technology on Surface Physics and Chemistry Laboratory, P.O. Box 9-35, Jiangyou 621908, China}
\date{\today}

%% abstract
%%%%%%%%%%%%%%%%%%%%%%%%%%%%%%%%%%%%%%%%%%%%%%%%%%%%%%%%%%%%%%%%%%%%%%%%%%

\begin{abstract}
The electronic structures of element cerium under high pressure remain unclear all the time. We tried to calculate the electronic structures of $\alpha'$, $\alpha''$, and $\epsilon$-Ce which only exist in the presence of pressure, by using the combination of traditional density functional theory and single-site dynamical mean-field theory. The momentum-resolved band structures, total and partial density of states, 4$f$ electronic configurations of these phases were exhaustively studied. We found that the 4$f$ electrons tend to be itinerant, and the hybridizations between the 4$f$ and $spd$ conduction electrons are remarkable. In addition, the fluctuations among the 4$f$ atomic eigenstates are prominent, especially for the $\epsilon$ phase, which leads to a slight modification of its 4$f$ occupancy. 
\end{abstract}

%% pacs number
%%%%%%%%%%%%%%%%%%%%%%%%%%%%%%%%%%%%%%%%%%%%%%%%%%%%%%%%%%%%%%%%%%%%%%%%%%

\pacs{71.10.-w, 71.27.+a, 71.30.+h, 74.20.Pq}
\maketitle

%% introduction
%%%%%%%%%%%%%%%%%%%%%%%%%%%%%%%%%%%%%%%%%%%%%%%%%%%%%%%%%%%%%%%%%%%%%%%%%%

\section{Introduction\label{sec:intro}} 

%% mysterious Ce
Ce is one of the most mysterious elements in the periodic table because of its unusual electronic structures which manifest in its multiple allotropic phases. Under ambient pressure, Ce may exhibit four allotropes ($\alpha$, $\beta$, $\gamma$, and $\delta$ phases), and undergo three successive solid phase transitions before reaching its liquid state~\cite{koskenmaki1978337}. When $P \sim 0.8$ GPa and at room temperature, strikingly, an iso-structural phase transition would take place between the $\alpha$ and $\gamma$ phases (both phases are in face-centered-cubic structure). Though the crystal structure is preserved during the transition, the atomic volume changes by $\sim$ 15\% which has not been observed in any other simple metals. The underlying physics and driving force of the $\alpha-\gamma$ phase transition have been, and are being to this day, warmly debated. Particularly, concerning the role played by the correlated 4$f$ electrons in the transition, it is still a matter of dispute and ongoing research~\cite{bj:1974,PhysRevLett.74.2335,PhysRevLett.49.1106,PhysRevB.46.5047,PhysRevLett.92.105702,PhysRevLett.101.165703,PhysRevB.89.184426}. 

\begin{figure}[t]
\centering
\includegraphics[width=\columnwidth]{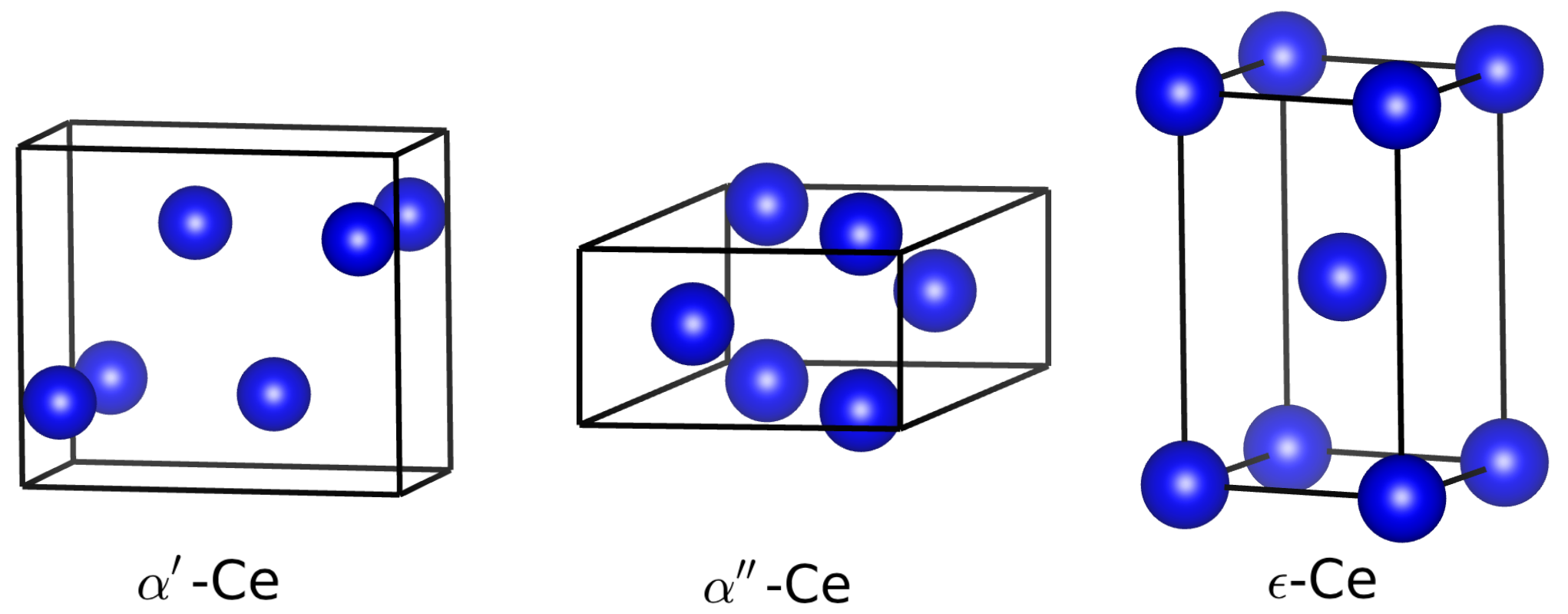}
\caption{(Color online). Schematic picture for the crystal structures of $\alpha'$, $\alpha''$, and $\epsilon$-Ce. \label{fig:struct}}
\end{figure}

%% the high pressure phase of Ce, basic information
Besides, under moderate pressure and temperature Ce will exhibit another three distinct phases, namely, the $\alpha'$, $\alpha''$, and $\epsilon$ phases (see Fig.~\ref{fig:struct}), which result in an extremely complicated $P-T$ phase diagram~\cite{alex:2012}. The crystal structure of the $\alpha'$ phase is orthorhombic (space group $Cmcm$). Noted that it is iso-structural with $\alpha$-U. The $\alpha''$ phase is monoclinic (space group $C_2/m$). The $\alpha$-Ce will transform into the two phases in the pressure range from 5 to 12 GPa, but which phase transition (i.e., $\alpha-\alpha'$ or $\alpha-\alpha''$ transition) could occur in a cerium sample strongly depends on the method of sample preparation. Upon further volume compression ($P > 12$ GPa) both the $\alpha'$ and $\alpha''$ phases can transform into the $\epsilon$ phase, which is body centered tetragonal (space group $I_4/mmm$). Recently, the inelastic X-ray scattering experiment has demonstrated that there exist strong electron-phonon coupling and pronounced phonon anomalies in the $\alpha'$ phase, which means that it is at the verge of lattice instability~\cite{PhysRevLett.108.045502}. Similar to $\alpha$-U, a charge density wave perhaps develops in the $\alpha'$-Ce. Furthermore, the superconductivity with $T_c \sim 2$ K found in the high-pressure phases of Ce can be attributed to the $\alpha'$ phase as well.

%% we know a little about the electronic structures of Ce under pressure
It is very difficult to prepare and identify pure high-pressure phases of Ce, which requires precise controls of pressure and thermal process~\cite{PhysRevLett.32.773}. Therefore, it is not easy to conduct extensively experimental researches. On the other hand, low-symmetry crystal structures, strong spin-orbit coupling, and notable electronic correlations in 4$f$ electrons also pose great challenges to the first-principles calculations~\cite{koskenmaki1978337}. As a result, though numerous efforts have been made to explore the basic properties of Ce under pressure, actually we know a little about the electronic structures of its high-pressure phases and there are a lot of puzzles that need to be solved. For instance, do the 4$f$ electrons remain itinerant or localized state~\cite{smith:198383}? What's the consequence in electronic structures since the interplay of spin-orbit coupling and 4$f$ electronic correlations? Are there strong hybridization between the 4$f$ and $spd$ bands as predicted by the Kondo volume collapse (KVC) scenario~\cite{PhysRevLett.49.1106,PhysRevB.46.5047}? Last, but by no means least, are the 4$f$ electronic configurations for these high-pressure phases the same, and are they similar to or different from those for the other phases under ambient pressure~\cite{PhysRevLett.94.036401,PhysRevB.67.075108,PhysRevLett.87.276404,PhysRevLett.87.276403,PhysRevB.72.115125,PhysRevLett.96.066402,PhysRevB.89.125113}? In order to answer the above questions, more experimental investigations and theoretical calculations are highly desired. 

%% purpose of this paper
Bearing these questions in mind, the purpose of this paper is to endeavour to build an unified picture for the evolution of electronic structures of high-pressure phases of Ce, by utilizing the state-of-the-art first-principles many-body approach. The rest of the paper is organized as follows. In Sec.~\ref{sec:method}, we briefly introduce the calculation method and parameters. In Sec.~\ref{sec:results}, we present all of the calculated results and make a detailed discussion. We also compare the $4f$ electronic configurations between the high-pressure phases and $\alpha$-Ce. Finally, Section~\ref{sec:summary} serves as a short summary and conclusion. 

%% methods
%%%%%%%%%%%%%%%%%%%%%%%%%%%%%%%%%%%%%%%%%%%%%%%%%%%%%%%%%%%%%%%%%%%%%%%%%%

\section{Methods\label{sec:method}} 

%% an overview of DFT + DMFT
In the present work, we employed the density functional theory in combination with the single-site dynamical mean-field theory (dubbed as DFT + DMFT) to study the electronic structures of $\alpha'$, $\alpha''$, and $\epsilon$-Ce. This approach can treat the strong $4f$ electronic correlations, spin-orbit coupling as well as on-site Kondo screening on the same footing, and incorporate them into realistic band structures based on DFT calculations. As for the basic principles and technical details of the DFT + DMFT method, please refer to the good reviews (see Refs.~[\onlinecite{RevModPhys.78.865}] and [\onlinecite{RevModPhys.68.13}]) and references therein. In fact, it is considered as the most powerful method established so far for the calculations of electronic structures of strongly correlated materials, and has been successfully applied to study many interesting problems in lanthanides and actinides, such as the best known $\alpha-\gamma$ phase transition in Ce~\cite{PhysRevLett.87.276403,PhysRevLett.87.276404,PhysRevLett.94.036401,PhysRevLett.96.066402,PhysRevB.89.125113,PhysRevB.67.075108,PhysRevB.72.115125}, the low-temperature hidden order phase of URu$_{2}$Si$_{2}$~\cite{RevModPhys.83.1301}, and the valence state fluctuations in Yb~\cite{PhysRevLett.102.246401} and Pu~\cite{shim:2007,Janoscheke:2015}. 

%% calculation details
We performed charge fully self-consistent DFT + DMFT calculations by using the \texttt{EDMFTF} software package~\cite{PhysRevB.81.195107}. The Kohn-Sham equation within the DFT part was solved by using the \texttt{WIEN2K} code, which implements a full-potential linear augmented plane-wave formalism~\cite{wien2k}. We selected the generalized gradient approximation to represent exchange-correlation functional~\cite{PhysRevLett.77.3865}. The hybridization expansion version continuous-time quantum Monte Carlo impurity solver (dubbed as CT-HYB)~\cite{RevModPhys.83.349,PhysRevB.75.155113,PhysRevLett.97.076405} was used to solve the quantum impurity model within the DMFT part. Undoubtedly, the 4$f$ bands were treated as correlated orbitals. The corresponding interaction parameters were on-site Coulomb interaction $U = 6.0$ eV and Hund's coupling $J = 0.7$ eV~\cite{PhysRevLett.87.276403,PhysRevLett.87.276404}. The experimental lattice structures and system temperature $T \sim 290$ K ($\beta = 40.0$) were used throughout the calculations.

%% results
%%%%%%%%%%%%%%%%%%%%%%%%%%%%%%%%%%%%%%%%%%%%%%%%%%%%%%%%%%%%%%%%%%%%%%%%%%

\section{Results\label{sec:results}}

%% gather figures here?
\begin{figure*}[ht]
\centering
\includegraphics[width=\textwidth]{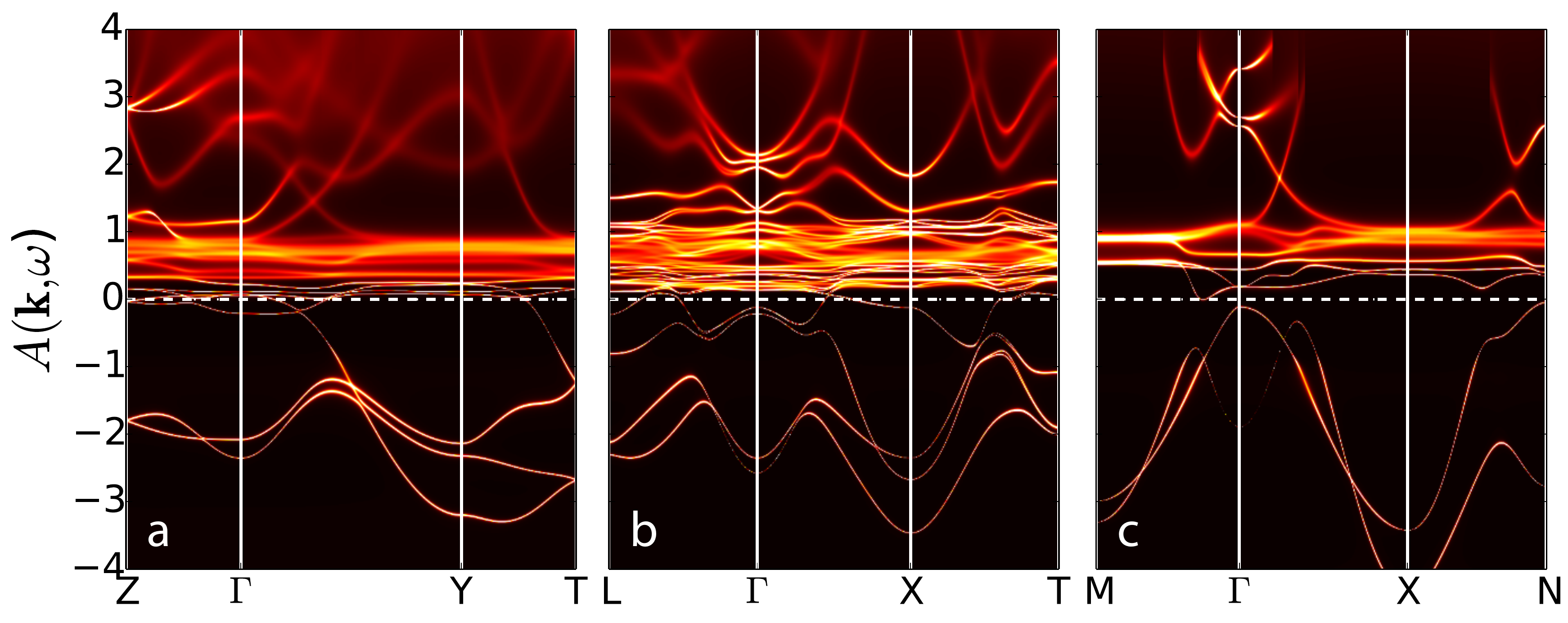}
\caption{(Color online). Momentum-resolved spectral functions $A(\mathbf{k},\omega)$ of Ce under high pressure. (a) $\alpha'$-Ce. (b) $\alpha''$-Ce. (c) $\epsilon$-Ce. They were calculated via the analytical continuation of Matsubara self-energy function~\cite{PhysRevB.81.195107}. The horizontal dashed lines represent the Fermi level. \label{fig:akw}}
\end{figure*}

\begin{figure*}[ht]
\centering
\includegraphics[width=\textwidth]{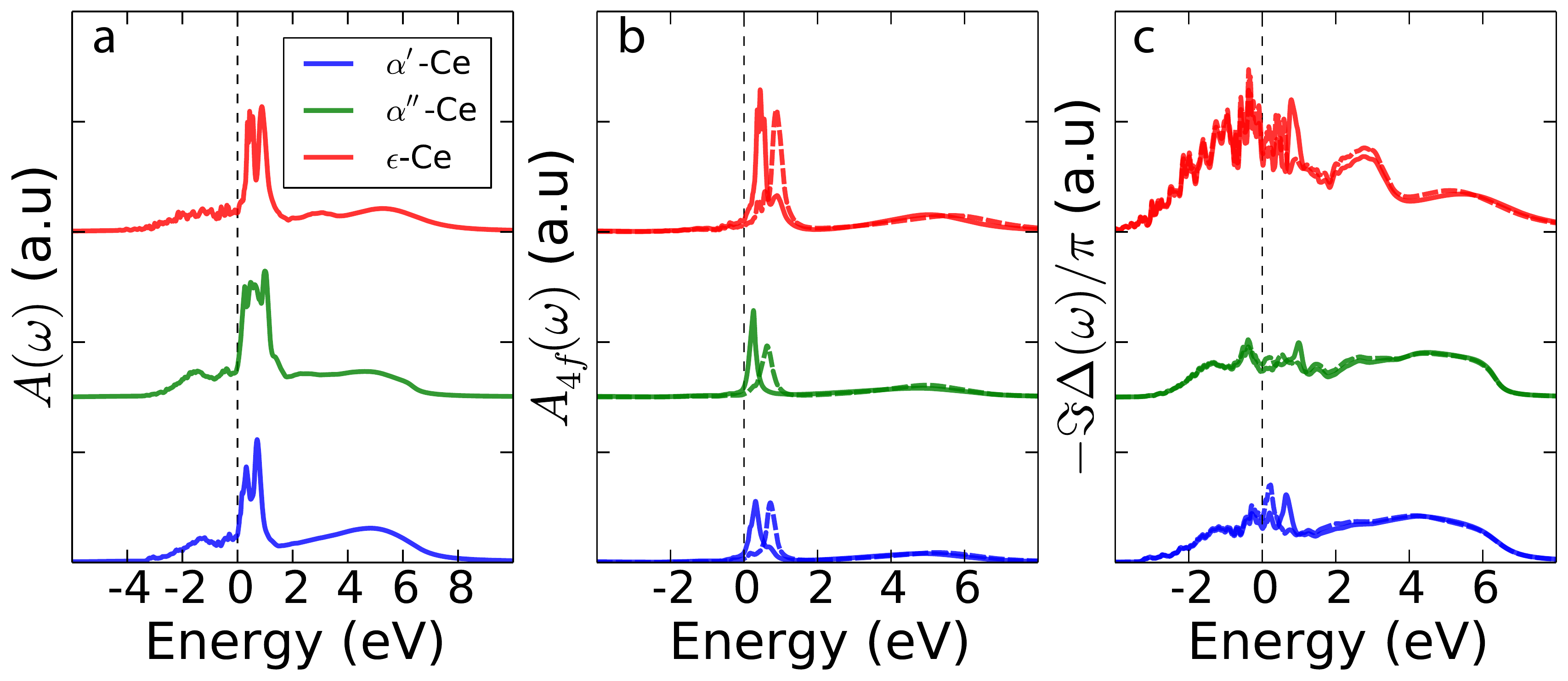}
\caption{(Color online). (a) Total density of states $A(\omega)$, (b) $4f$ partial density of states $A_{4f}(\omega)$, and (c) 4$f$ hybridization functions $-\Im \Delta(\omega)/\pi$ of Ce under high pressure. They were calculated via the analytical continuation of Matsubara self-energy function~\cite{PhysRevB.81.195107}. The data are rescaled and shifted vertically for a better view. In panels (b) and (c), the spectra for the $4f_{5/2}$ and $4f_{7/2}$ components are plotted as solid and dashed lines, respectively. The vertical dashed lines represent the Fermi level. \label{fig:dos}}
\end{figure*}

\begin{figure*}[ht]
\centering
\includegraphics[width=\textwidth]{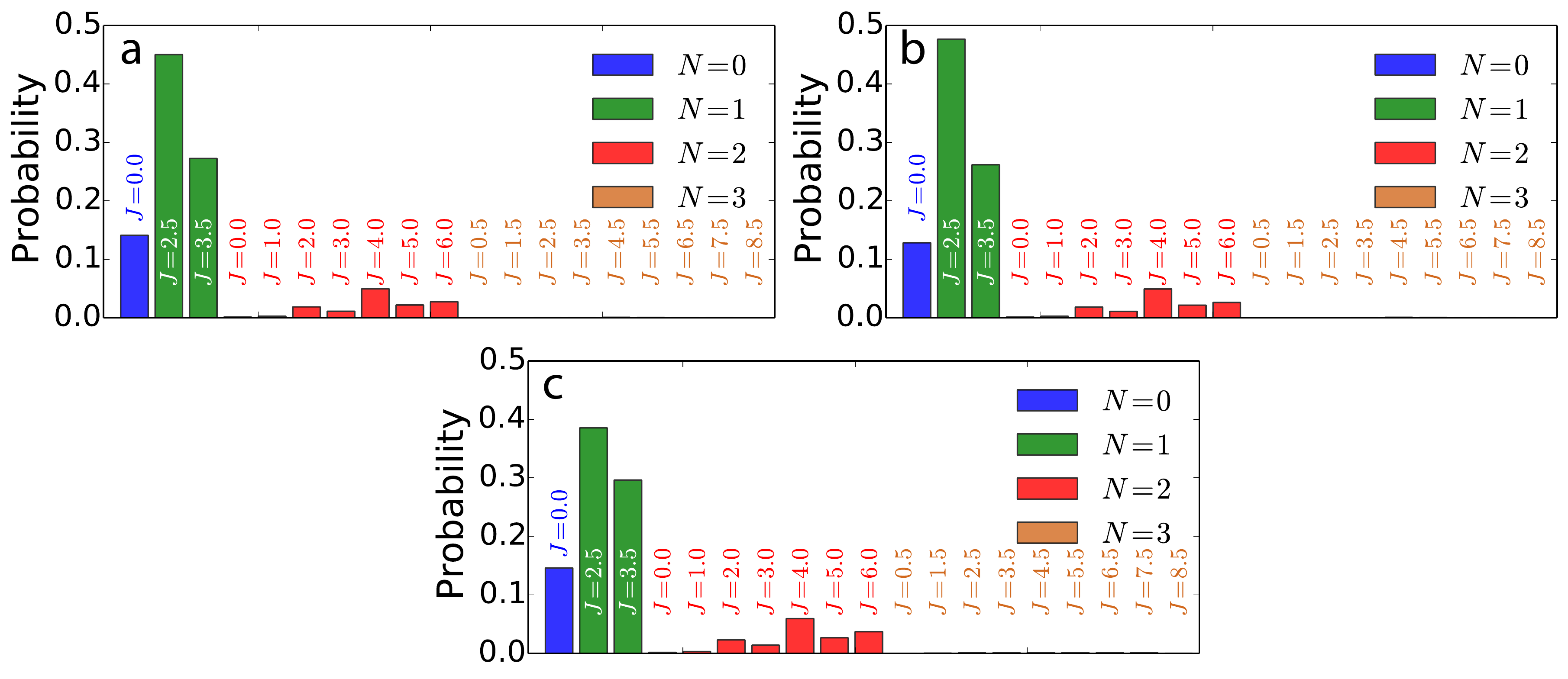}
\caption{(Color online). Probabilities of the 4$f$ atomic eigenstates (or equivalently valence state histograms) of Ce under high pressure. (a) $\alpha'$-Ce. (b) $\alpha''$-Ce. (c) $\epsilon$-Ce. Note that the histograms from the $\bm{N} = 3$ atomic eigenstates are too trivial to be seen in this figure. \label{fig:prob1}}
\end{figure*}

\begin{figure}[ht]
\centering
\includegraphics[width=\columnwidth]{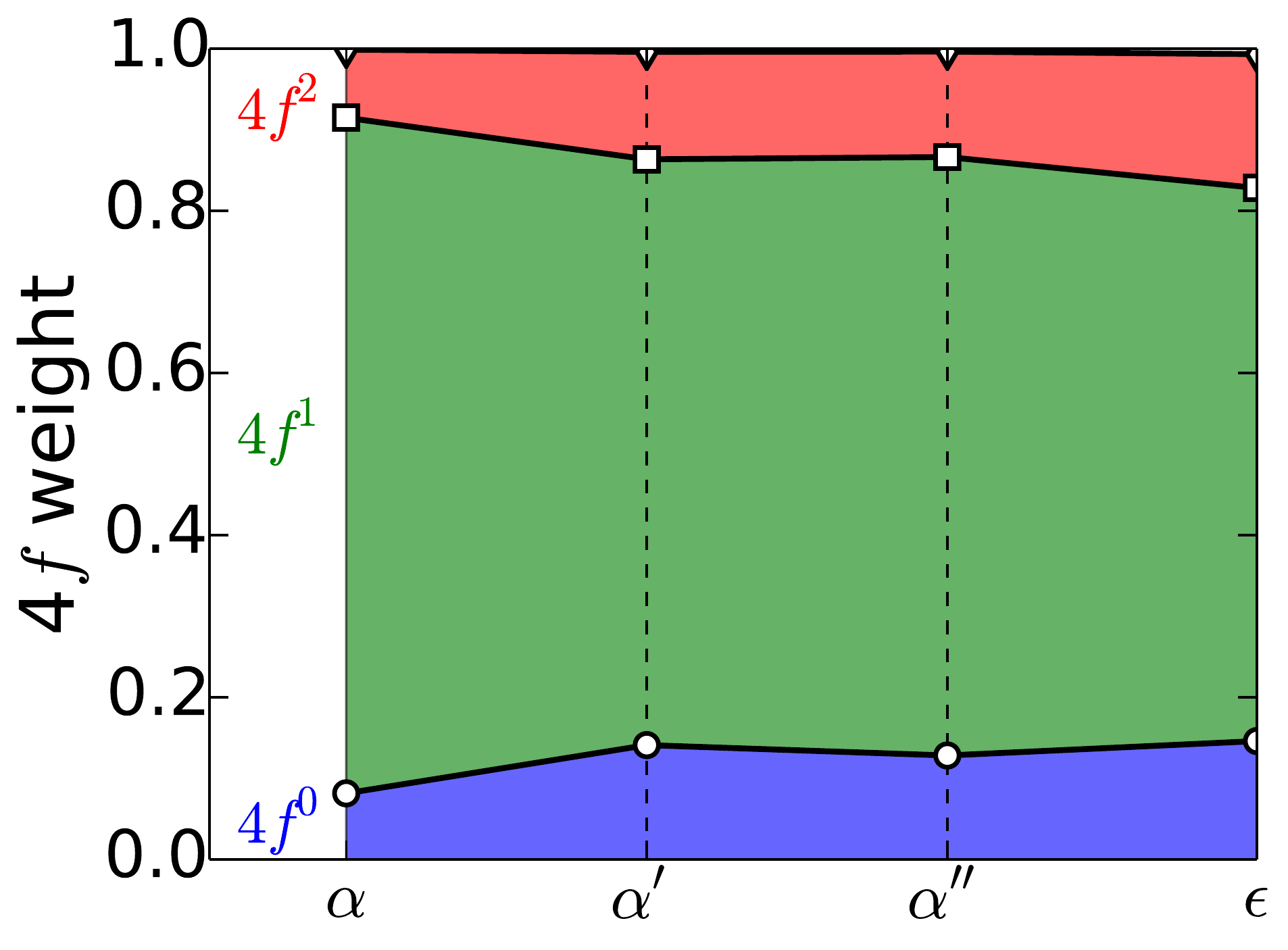}
\caption{(Color online). Distributions of 4$f$ electronic configurations of $\alpha$,  $\alpha'$, $\alpha''$, and $\epsilon$-Ce. \label{fig:prob2}}
\end{figure}

\emph{Momentum-resolved spectral functions $A(\mathbf{k},\omega)$.} The calculated band structures of the three high-pressure phases along the high-symmetry lines in the Brillouin zone are displayed in Fig.~\ref{fig:akw}. Since most of the 4$f$ bands in Ce are unoccupied, their positions are well above the Fermi level. As $E_{F} < \omega < 1.5$ eV, we observe some intense and stripe-like features, which are mainly associated with the 4$f$ bands. Their bandwidths are approximately $0.8 \sim 1.2$ eV, which are much smaller than the Coulomb interaction strength (i.e., $W \ll U$). It is an exact evidence to support the statement that the 4$f$ bands in the high-pressure phases of Ce are still strongly correlated~\cite{RevModPhys.68.13}. In addition, we also observe heavy $c-f$ hybridizations between the 4$f$ and the $spd$ conduction bands in this energy window. As $\omega > 1.5$ eV and $\omega < E_{F}$, the general features become somewhat blurry. The bands in this energy range exhibit remarkable dispersions and large bandwidths. Apparently, they belong to the $spd$ itinerant bands to a large extent. 

Now let us concentrate on the similarities and differences in the band structures of the three phases. As for the $\alpha'$ and $\alpha''$ phases, since their lattice volumes are very close and their crystal structures are tightly related, the corresponding band structures are quite similar and share some common features, such as the positions of the 4$f$ and valence bands. However, for the $\epsilon$ phase, which exists under higher pressure and thus has a smaller lattice volume~\cite{alex:2012}, the center of the correlated 4$f$ bands shifts upwards slightly and the itinerant $spd$ bands exhibit much more significant dispersions. Finally, from Fig.~\ref{fig:akw}, we can speculate that for the $\epsilon$ phase the volume encapsulated by the Fermi surface is the smallest. In the $\alpha'$ and $\alpha''$ phases, the Fermi surfaces that enclose the $\Gamma$ point are hole-type and electron-type, respectively. 

\emph{Density of states $A(\omega)$.} The total density of states $A(\omega)$ are shown in Fig.~\ref{fig:dos}(a). In the vicinity of the Fermi level, there is the so-called Kondo resonance peak with sharp two-peak structure. This splitting is owing to the spin-orbit coupling effect. Away from the Fermi level, the spectrum spreads out and becomes almost featureless. Just as discussed above, the spectra for the $\alpha'$ and $\alpha''$ phases are very similar, including the magnitude of the Kondo resonance peak, the two satellite peaks between -2 eV and $E_{F}$, and the total bandwidth. As for the $\epsilon$ phase, its spectrum turns broader and has a stronger Kondo resonance peak. 

The 4$f$ partial density of states $A_{4f}(\omega)$ are shown in Fig.~\ref{fig:dos}(b). Clearly, according to the difference between $A(\omega)$ and $A_{4f}(\omega)$, the contributions from the 4$f$ bands are predominant near the Fermi level. They compose the Kondo resonance peak. On the other hand, below the Fermi level, the itinerant $spd$ bands are the most important (the 4$f$ spectral weight approaches zero). Due to the spin-orbit coupling, the 4$f$ bands are split into two components, namely $4f_{5/2}$ and $4f_{7/2}$. The $4f_{5/2}$ component has a lower energy and is closer to the Fermi level. The lower Hubbard bands are almost invisible since the 4$f$ occupancy is small. The upper Hubbard bands look like broad ``humps'', and they locate roughly from 2 eV to 8 eV. All these features are comparable to those already found in the $\alpha$ phase by using the DFT + DMFT method~\cite{PhysRevB.72.115125,PhysRevLett.87.276403,PhysRevLett.87.276404,PhysRevB.67.075108} and photoemission spectroscopy~\cite{PhysRevB.29.3028,PhysRevB.28.7354}. Compared to the $\alpha'$ and $\alpha''$ phases, the $A_{4f}(\omega)$ of the $\epsilon$ phase is a bit different. The peaks ascribed to the $4f_{5/2}$ and $4f_{7/2}$ components (and the upper Hubbard bands) have larger intensity, and their positions are shifted to high-energy regime marginally.

In Fig.~\ref{fig:dos}(c), we visualize the imaginary parts of the real-frequency hybridization functions, i.e., $-\Im \Delta(\omega)/\pi$, which are approximately proportional to the strength of hybridization effect between the strongly correlated 4$f$ bands and the weakly correlated (or non-correlated) $spd$ bands. We can see that the hybridization effect is very strong, especially near the Fermi level. It indicates that the KVC scenario not only can be used to explain the $\alpha-\gamma$ phase transition~\cite{PhysRevLett.49.1106,PhysRevB.46.5047}, but also is useful for understanding the underlying electronic structures of the high-pressure phases of Ce. Noticed that the hybridization function of the $\epsilon$ phase is much higher than the others, which implies that its 4$f$ electrons tend to be more itinerant. Naturally, we believe that the valence state fluctuation in the $\epsilon$ phase must be the most pronounced among cerium's allotropes. We will discuss this issue later.  

\emph{Distribution of atomic eigenstates.} The CT-HYB quantum impurity solver is capable of computing the valence state histogram $p_{\Gamma}$, which means the probability to find a valence state electron (such as the $4f$ electrons in Ce) in a given atomic eigenstate $|\psi \rangle$ (labelled by $\Gamma$)~\cite{PhysRevB.75.155113}. It provides us a powerful lens to learn the subtle electronic structures of strongly correlated systems directly. Besides the first successful application for the 4$f$ electrons in $\alpha$ and $\gamma$-Ce, it has been adopted to analyze the status of 5$f$ electrons in $\alpha$ and $\delta$-Pu~\cite{shim:2007,Janoscheke:2015}, and the 4$f$ localized-itinerant crossover in heavy fermion compound CeIn$_{3}$~\cite{PhysRevB.94.075132}. Here, we would like to utilize it to unveil whether the 4$f$ electrons in the high-pressure phases of Ce are localized or not. In the presence of spin-orbit coupling, the 4$f$ atomic eigenstates could be classified by using the good quantum numbers $\bm{N}$ (total occupancy) and $\bm{J}$ (total angular momentum). In the present work, in order to accelerate the calculations, we made a crucial truncation that only those atomic eigenstates whose $\bm{N} \in [0,3]$ were taken into considerations~\cite{PhysRevB.75.155113}. Next we will prove that the contributions from the atomic eigenstates with larger occupancy ($\bm{N} \geq 3$) are trivial and the truncation we used is reasonable.  

The calculated valence state histograms are shown in Fig.~\ref{fig:prob1}. As is seen in this figure, the predominant atomic eigenstates are $| \bm{N} = 1, \bm{J} = 2.5\rangle$ and $| \bm{N} = 1, \bm{J} = 3.5\rangle$. Their probabilities add up to about 68\% $\sim$ 74\%. The less important atomic eigenstate is $| \bm{N} = 0, \bm{J} = 0.0 \rangle$, which amounts to about 13\% $\sim$ 15\%. For the $\bm{N} = 2$ case, the contributions from various atomic eigenstates account for about 13\% $\sim$ 17\%. They are comparable and no dominant ones. The probabilities for the atomic eigenstates with $\bm{N} = 3$ are tiny ($< 1\%$ in total), so it is hardly to see them in Fig.~\ref{fig:prob1}. Since the contributions from the $\bm{N} \ge 3$ atomic eigenstates are unimportant, it is acceptable to discard them to improve the efficiency during the DFT + DMFT calculations. Actually, we rerun all of the calculations with $\bm{N} \in [0,2]$ and $\bm{N} \in [0,4]$. The calculated results are practically identical.

Next, let us compare the valence state histograms of the three phases. First, in the $\alpha'$ and $\alpha''$ phases, the probability of the $|\bm{N} = 1, \bm{J} = 2.5 \rangle$ eigenstate is nearly twice as much as the one of the $|\bm{N} = 1, \bm{J} = 3.5 \rangle$ eigenstate. While for the $\epsilon$ phase, the probabilities for the two atomic eigenstates are close. Second, the probabilities for the $\bm{N} = 0$, 2, and 3 cases in the $\epsilon$ phase are somewhat larger than those in the $\alpha'$ and $\alpha''$ phases. All these facts reveal that the 4$f$ electrons in the high-pressure phases of cerium are not always bound to the $\bm{N} = 1$ case, and tend to hop among a variety of atomic eigenstates with different occupancy and angular momentum. Especially for the $\epsilon$ phase, this tendency is the most noticeable.

\emph{Valence state fluctuations.} From the 4$f$ valence state histograms, we can easily evaluate the distributions of 4$f$ electronic configurations, and then discuss the valence state fluctuation phenomenon which usually manifests itself in the rare earth compounds (such as the Ce-based, Sm-based, and Yb-based heavy fermion systems~\cite{PhysRevB.87.115107,PhysRevLett.102.246401}) and the actinides (such as Pu~\cite{Janoscheke:2015,shim:2007}). Supposed that the weight for the $4f^{i}$ (where $i \in [0,3]$) configuration is $\mathcal{W}(4f^{i})$, then it can be calculated via the following equation,
\begin{equation}
\mathcal{W}(4f^{i}) = \sum_{\bm{N}}\sum_{\bm{J}} \delta(\bm{N} - i) p_{\Gamma}. 
\end{equation}
We tried to calculate the weights for the $4f^{[0 - 3]}$ electronic configurations. The results are presented in Fig.~\ref{fig:prob2}. In order to gain a clear impression about the evolution of electronic configurations, we performed additional calculations for the $\alpha$ phase, and supplemented its data to this figure. Obviously, for $\mathcal{W}(4f^{1})$, we have $\alpha > \alpha' \approx \alpha'' > \epsilon $. However, for $\mathcal{W}(4f^{0})$ or $\mathcal{W}(4f^{2})$, the sequence is inverse, i.e., $\alpha < \alpha' \approx \alpha'' < \epsilon $.

Based on the obtained data, we reached the following conclusions. First, the distributions of electronic configurations for the $\alpha'$ and $\alpha''$ phases are similar in all respects. Second, in the high-pressure phases of cerium, the 4$f$ valence state fluctuations are more significant than the phases under ambient pressure. Particularly, in the $\epsilon$ phase, the 4$f$ valence state fluctuation is much stronger than the other allotropes. Third, strong valence state fluctuation will modify the valence state occupancy inevitably. Thus we employed the following equation to evaluate the averaged 4$f$ occupancy $N_{4f}$~\cite{PhysRevB.94.075132},
\begin{equation}
N_{4f} \approx \sum^{3}_{i = 0} \mathcal{W}(4f^{i}) \times i.
\end{equation}
The calculations suggest that the $N_{4f}$ for the $\alpha'$, $\alpha''$, and $\epsilon$ phases are 0.98, 0.99, and 1.03, respectively~\cite{occupy}. Alternatively, the 4$f$ occupancy is only modified slightly in this case. 

\begin{figure}[ht]
\centering
\includegraphics[width=\columnwidth]{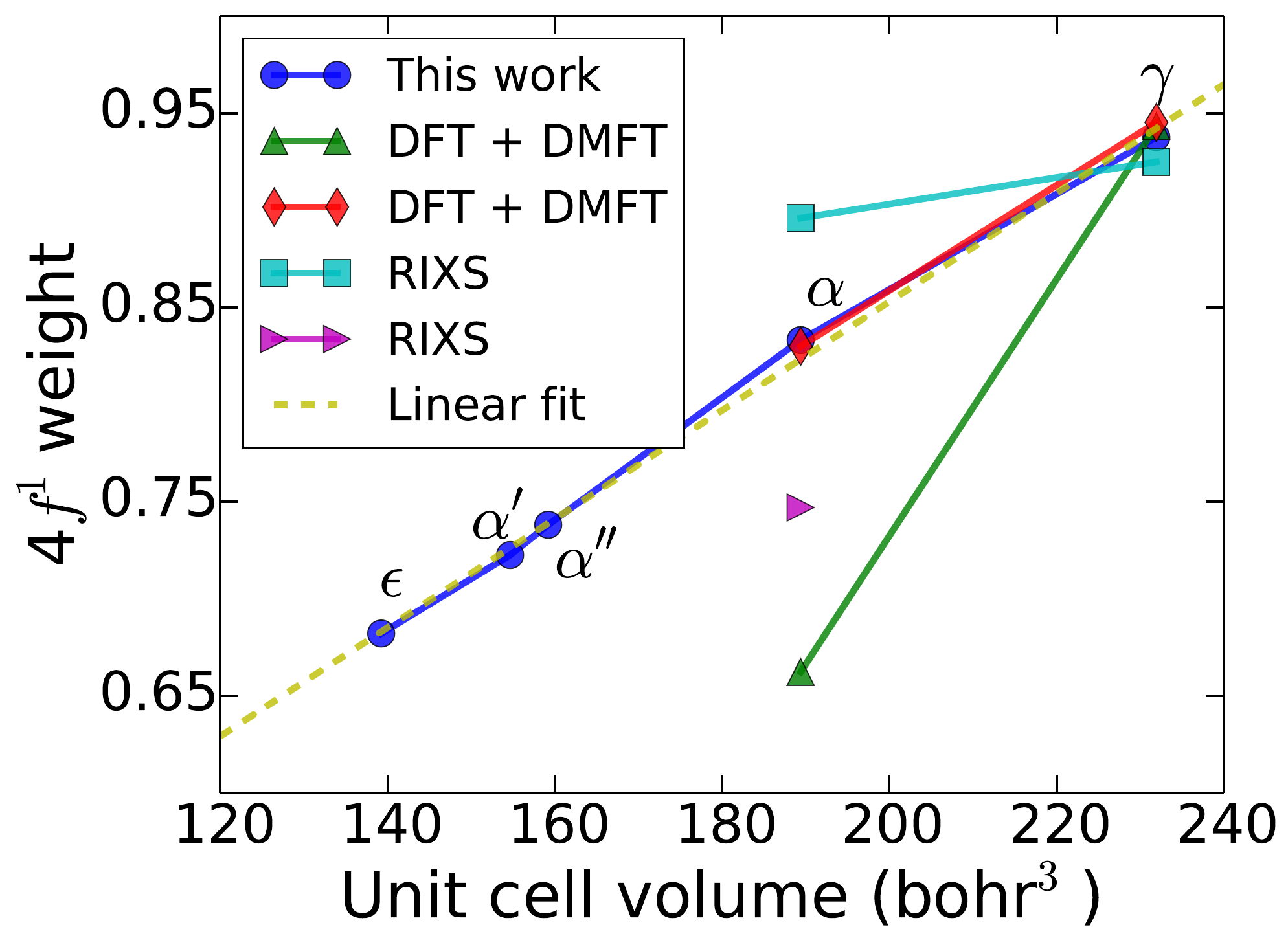}
\caption{(Color online). The 4$f^{1}$ weight $\mathcal{W}(4f^{1})$ with respect to the unit cell volumes $V$~\cite{volume} of various phases of Ce. The DFT + DMFT data for reference are extracted from Reference~\cite{PhysRevB.75.155113} (upper triangle symbols) and \cite{PhysRevLett.87.276403} (diamond symbols). The resonant inelastic X-Ray scattering (RIXS) experimental data for $\alpha$-Ce and $\gamma$-Ce are taken from Reference~\cite{PhysRevLett.96.237403}. For $\alpha$-Ce, the data were measured under 10 kbar (square symbol) and 20 kbar (right triangle symbol). For $\gamma$-Ce, the data were measured under 0 kbar (square symbol). See main text for the detailed explanations. \label{fig:vol}}
\end{figure}

\emph{Discussions.} Under external pressure, the unit cell volumes of cerium is reduced, and the corresponding electronic structure (including the band structure, density of states, Fermi surface, valence state histogram, and distribution of electronic configuration) is tuned gradually. Here, we would like to figure out the relationship between the lattice volume and the 4$f$ electronic configurations.  

In Fig.~\ref{fig:vol}, we try to plot the $\mathcal{W}(4f^{1})$ against lattice volume~\cite{volume}. Besides the three high-pressure phases, the available data for the $\alpha$ and $\gamma$ phases are also included. Surprisingly, we find that the $\mathcal{W}(4f^{1}) - V$ curve exhibits a quasi-linear behavior. Then we used the formula $f(x) = ax + b$ to fit the data (please see the dashed dark-yellow line in Fig.~\ref{fig:vol}). The fitting parameters $a$ and $b$ are 0.00279767 and 0.293449, respectively. This linear behavior is easily to be understood. The larger the lattice volume is, the more localized the 4$f$ electrons are, and the smaller the 4$f$ valence state fluctuation will be. As a result, the 4$f$ electrons have a tendency to stay at the $4f^{1}$ electronic configuration instead of hopping to and fro, with respect to the increment of lattice volume. On the other hand, we think that the $\mathcal{W}(4f^{1})$ could be considered as a quantitative tool to measure the valence state fluctuation and electronic localized degree of freedom for Ce-based system.  

\section{summary\label{sec:summary}} 

In conclusion, we studied the electronic structures of three high-pressure phases of Ce by using the charge fully self-consistent DFT + DMFT method. We found that the 4$f$ electrons exhibit more itinerant features. And the valence state fluctuation in the $\epsilon$ phase is the most remarkable, which is precisely described by the quantity $\mathcal{W}(4f^{1})$. To the best of our knowledge, it is the first time to adopt the first-principles many-body approach to study the $\alpha'$, $\alpha''$, and $\epsilon$-Ce. It is also the first step to build an unified picture for the electronic structures and related physical properties of all allotropes of Ce. Unfortunately, the corresponding experiments are rare in the literatures. Our results serve as critical predictions, and require further experimental examinations. 

\begin{acknowledgments}
This work was supported by the Natural Science Foundation of China (No.~11504340), the Foundation of President of China Academy of Engineering Physics (No.~YZ2015012), the Discipline Development Fund Project of Science and Technology on Surface Physics and Chemistry Laboratory (No.~201502), and the Science Challenge Project of China (No.~JCKY2016212A56401).
\end{acknowledgments}

%% reference
%%%%%%%%%%%%%%%%%%%%%%%%%%%%%%%%%%%%%%%%%%%%%%%%%%%%%%%%%%%%%%%%%%%%%%%%%%

\bibliography{ce}

\begin{thebibliography}{38}
\bibitem [{occ()}]{occupy}%
  \BibitemOpen
  \href@noop {} {}\bibinfo {note} {We can evaluate the 4$f$ occupations
  $N_{4f}$ through the impurity Green's function $G(\tau)$ or $G(i\omega_n)$,
  but usually the calculated values are a little smaller.}\BibitemShut {Stop}%
\end{thebibliography}

\end{document}